\documentclass{pnastwo}

\usepackage[dvips]{graphicx}

\usepackage{longtable}%
\usepackage{bm}%
\usepackage{graphicx}
\usepackage{float}
\usepackage[T1]{fontenc}
\usepackage{amsmath}
\usepackage{amssymb}
\usepackage[pdftex]{color}

\newcommand{\<}{\langle}
\renewcommand{\>}{\rangle}

\newcommand{\IRS}{I_{\text{RS}}} 
\newcommand{\ICS}{I_{\text{CS}}}
\newcommand{\TRS}{T_{\text{RS}}} 
\newcommand{\TCS}{T_{\text{CS}}}
\newcommand{\lRS}{\lambda_{\text{RS}}} 
\newcommand{\lCS}{\lambda_{\text{CS}}}

\newcommand{\bx}{\mathbf{x}}
\newcommand{\by}{\mathbf{y}}
\newcommand{\bPhi}{\mathbf{\Phi}}

\usepackage{PNASTWOF}

\url{www.pnas.org/cgi/doi/10.1073/pnas.0709640104}
\copyrightyear{2008}
\issuedate{Issue Date}
\volume{Volume}
\issuenumber{Issue Number}
\begin{document}

\title{Compressive Fluorescence Microscopy \\ for Biological and Hyperspectral Imaging}

\author{Vincent Studer\affil{1}{Univ. Bordeaux, Interdisciplinary Institute for Neuroscience, UMR 5297, F-33000, Bordeaux, France.}\affil{2}{CNRS, Interdisciplinary Institute for Neuroscience, UMR 5297, F-33000, Bordeaux, France.}, J\'erome Bobin\affil{3}{CEA Saclay, IRFU/SEDI-Sap},
  Makhlad Chahid\affil{1}{}\affil{2}{},S. Hamed Shams Mousavi\affil{1}{}\affil{2}{}, Emmanuel
  Candes\affil{4}{Departments of Mathematics, of Statistics and of
    Electrical Engineering, Stanford University} \and Maxime
  Dahan\affil{5}{Laboratoire Kastler Brossel, CNRS UMR 8552, \'Ecole
    Normale Sup\'erieure, Universit\'e Pierre et Marie Curie-Paris 6}}

\contributor{Submitted to Proceedings of the National Academy of Sciences
of the United States of America}

\maketitle

\begin{article}
\begin{abstract} 
  The mathematical theory of compressed sensing (CS) asserts that one
  can acquire signals from measurements whose rate is much lower than
  the total bandwidth. Whereas the CS theory is now well developed,
  challenges concerning hardware implementations of CS-based
  acquisition devices---especially in optics---have only started being
  addressed. This paper presents an implementation of compressive
  sensing in fluorescence microscopy and its applications to
  biomedical imaging. Our CS microscope combines a dynamic structured
  wide-field illumination and a fast and sensitive single-point
  fluorescence detection to enable reconstructions of images of
  fluorescent beads, cells and tissues with undersampling ratios
  (between the number of pixels and number of measurements) up to
  32. We further demonstrate a hyperspectral mode and record images
  with 128 spectral channels and undersampling ratios up to 64,
  illustrating the potential benefits of CS acquisition for
  higher dimensional signals which typically exhibits extreme
  redundancy.  Altogether, our results emphasize the interest of
  CS schemes for acquisition at a significantly reduced rate and point out to some
  remaining challenges for CS fluorescence microscopy. 
  \end{abstract}

%% When adding keywords, separate each term with a straight line: |
\keywords{compressive sensing | sparse images | fluorescence microscopy | biological imaging}

%% Optional for entering abbreviations, separate the abbreviation from
%% its definition with a comma, separate each pair with a semicolon:
%% for example:
%% \abbreviations{SAM, self-assembled monolayer; OTS,
%% octadecyltrichlorosilane}

 \abbreviations{CS, compressed sensing; CFM, compressive fluorescence microscopy}
 
\section{Introduction}
\label{sec:intro}
%% The first letter of the article should be drop cap: \dropcap{}
\dropcap{F}luorescence microscopy is a fundamental tool in basic and
applied biomedical research. Because of its optical sensitivity and
molecular specificity, it is employed in an increasing number of
applications which, in turn, are continuously driving the development
of advanced microscopy systems that provide imaging data with ever
higher spatio-temporal resolution and multiplexing capabilities. In
fluorescence microscopy, one can schematically distinguish two kinds
of imaging approaches, differing by their excitation and detection
modalities \cite{Mertz}.  In wide-field (WF) microscopy, a large
sample area is illuminated and the emitted light is recorded on a
multi-detector array, such as a CCD camera. In contrast, in raster
scan (RS) microscopy, a point excitation is scanned through the sample
and a point detector is used to detect the fluorescence signal at each
position.

While very distinct in their implementation and applications, these
imaging modalities have in common that the acquisition is independent
of the information content of the image. Rather, the number of
measurements, either serial in RS or parallel in WF, is imposed by the
Nyquist-Shannon theorem. This theorem states that the sampling
frequency (namely the inverse of the image pixel size) must be twice
the bandwidth of the signal, which is determined by the diffraction
limit of the microscope lens equal to $\lambda/2\text{NA}$ ($\lambda$
is the optical wavelength and $\text{NA}$ the objective numerical
aperture).  Yet, most images, including those of biological interest,
can be described by a number of parameters much lower than the total
number of pixels. In everyday's world, a striking consequence of this
compressibility is the ability of consumer cameras with several
megapixel detectors to routinely reduce the number of bits in a raw
data file by an order of magnitude or two without substantial
information loss. To quote from David Brady: ``if it is possible to
compress measured data, one might argue that too many measurements
were taken'' \cite{Brady}.

%A recent line of research going under the name of {\em compressed} or
%{\em compressive sensing} (CS) has addressed what seems to be

% However, it is known that, in practice, many images are
% compressible (or, equivalently here, sparse), meaning that they depend
% on a number of degrees of freedom $K$ that is smaller that their size
% $N$. The difference between $K$ and $N$ can be significant, which is
% why cameras with several megapixel detectors produce images that can
% be stored in relatively small files. However, the compression remains
% a post-acquisition processing step.

The recent mathematical theory of {\em compressed} or {\em compressive
  sensing} (CS -- see \cite{Donoho06,CRT06}) has addressed this
challenge and shown how the sensing modality could be modified to
reduce the sampling rate of objects which are sparse in the sense that
their information content is lower than the total bandwidth or the
number of pixels suggest. The fact that one can sample such signals
non-adaptively and without much information loss---if any at all---at
a rate close to the image information content (instead of the total
bandwidth) has important consequences, especially in applications
where sensing modalities are slow or costly. To be sure, the
applications of CS theory to data acquisition are rapidly growing in
fields as diverse as medical resonance imaging
\cite{MRI_Lustig,Trzasko}, analog-to-digital conversion
\cite{RMPI,Xampling} or astronomy \cite{CSAstro}. 

In optics, the interest in CS has been originally spurred by the
demonstration of the so-called ``single-pixel camera''
\cite{Baraniuk}. Since then, reports have explored the potential of CS
for visible and infrared imaging \cite{Brady06,Segev}, holography
\cite{Brady09} or ghost imaging \cite{Ghost}. In microscopy, the
feasibility of CS measurements has recently been demonstrated
\cite{Wu2011}. Altogether, these results open exciting prospects,
notably for the important case of biomedical imaging. Having said
this, there are very few results about the performance of CS hardware
devices on relevant biological samples. As such samples often have low
fluorescence, it is especially important to understand how the
associated noise will affect the acquisition and reconstruction
schemes.

In this paper, we describe Compressive Fluorescence Microscopy (CFM),
a novel modality for fluorescence biological and hyperspectral imaging
based on the concepts of CS theory. In CFM, the sample is excited with
a patterned illumination and its fluorescence is collected on a point
detector. Images are computationally reconstructed from measurements
corresponding to a set of appropriately chosen patterns. Therefore, CFM benefits
from many advantages associated with RS techniques, namely, high dynamic
range, facilitated multiplexing, and wide spectral range (from the UV
to the IR). In truth, the benefits of CS are particularly appealing in
biology where fast, high-resolution and multicolor imaging is highly
sought after.

The paper is organized as follows.  We begin by recalling the
principles of CS theory for optical imaging. We then turn to the
description of the practical implementation of CFM and of the sensing
protocol. Our techniques are subsequently applied to image several
relevant samples, including fluorescent beads, cultured cells and
tissues.  By extending our implementation, we further demonstrate the
possibility of hyperspectral acquisition with up to 128 different
spectral channels. A final contribution is a careful study of various
noise trade-offs for CFM. We conclude the paper with a discussion of
prospective CFM developments.

%%%%%%%%%%%%%%%%%%%%%%%%%%%%%%%%%%%%%%%%%%%%%%%
\section{Compressed Sensing Framework}
\label{sec:cs}
%%%%%%%%%%%%%%%%%%%%%%%%%%%%%%%%%%%%%%%%%%%%%%%

We wish to image a two-dimensional sample $\mathbf{x} = \{x[i]\}$, a
distribution of fluorescent probes, in which $x[i]$ is the value of
$\mathbf{x}$ at the pixel/location $i$ (thus one can view pixel intensities
$x[i]$ as the coefficients of the image $\mathbf{x}$ in a basis of localized
functions, namely, the Dirac basis). We represent this object in a
basis $\mathbf{W}$ of our choosing and write
\[
\mathbf{x} = \sum_{p} c[p] \mathbf{w_p} = \mathbf{W}\mathbf{c}, 
\]
where the $\mathbf{w_p}$'s are (orthogonal) basis functions and the
$c[p]$'s are the coefficients of $\mathbf{x}$ in the expansion. We say
that the signal is $K$-sparse if at most $K$ of these coefficients are
nonzero. An empirical fact is that most images of interest are well
approximated by $K$-sparse expansions with $K$ much less than the
number of pixels $N$, and this is the reason why data compression is
effective; one can store and transmit quantizations of the large
coefficients, ignore the small ones, and suffer little distortion.

In our imaging setup, we measure correlations between the image of
interest $\mathbf{x}$ and sensing waveforms $\mathbf{\phi_k}$ taken
from another basis $\mathbf{\Phi}$; that is, we measure
\begin{equation}
\label{eq:data}
y_k = \<\mathbf{x},\mathbf{\phi_k}\> = \sum_i x[i] \phi_k[i].
\end{equation}
Here, $\mathbf{\phi_k}$ is an illumination or intensity pattern so
that $y_k$ is obtained by collecting all the fluorescence
corresponding to those pixels that have been illuminated on a
single-point detector. Wide-field and point-like excitation are two
extreme cases, corresponding respectively to a uniform sensing
waveform ($\phi_k[i]=1$ for all $i$) and to a spike or Dirac waveform.

In its simplest form, CS theory asserts that if the signal
$\mathbf{x}$ is sparse in the representation $\mathbf{W}$, then only
few measurements of the form \eqref{eq:data} are sufficient for
perfect recovery provided the sensing and representation waveforms,
respectively $\mathbf{\phi_k}$ and $\mathbf{w_p}$, are {\em incoherent}
\cite{CRT06,CR07}. Two systems are said to be incoherent if any
element in one of the two cannot be expressed as a sparse linear
combination of elements taken from the other. Formally, the coherence
between two orthobases $\mathbf{W}$ and $\mathbf{\Phi}$ of
$\mathbb{R}^N$ is measured by the parameter
$\mu(\mathbf{W},\mathbf{\Phi})$ ranging between $1$ and $N$:
\begin{equation}
\label{eq:coherence}
\mu(\mathbf{W},\mathbf{\Phi}) = N \, \max_{p,k} |\langle \mathbf{w_p} , \mathbf{\phi_k} \rangle|^2.
\end{equation}

The Fourier and Dirac bases are in this sense maximally incoherent (we
need many spikes to synthesize a sinusoid and vice versa) and $\mu =
1$.  On the opposite, two identical bases are maximally coherent and,
in this case, $\mu = N$. Hence, incoherence expresses the idea of the
level of dissimilarity between any two representations of a
signal. With this in mind, one perceives how each incoherent
measurement---a projection on an element of the basis
$\mathbf{\Phi}$---conveys a little bit of information about all the
entries of the coefficient vector $\mathbf{c}$. An important result in
CS theory states that $K$-sparse signals can be recovered exactly from
comparably few measurements in an incoherent system. Further, recovery
is achieved by solving a tractable optimization program---a linear
program. One solves
\begin{equation}
  \min_{\mathbf{c} \in \mathbb{R}^N} \|\mathbf{c}\|_{\ell_1} \text{ subject to } y_k = \langle \mathbf{\phi_k} , \mathbf{W}\mathbf{c}\rangle, \text{ for all } k= 1 \ldots M.
\label{program}
\end{equation}
When $M$ measurements are chosen uniformly at random from the basis
$\mathbf{\Phi}$, the recovery is exact with very high probability;
that is, the solution sequence $\hat {\mathbf{c}}$ obeys $ \hat
{\mathbf{x}} = \sum_p \hat{c}[p] \mathbf{w_p} = \mathbf{x}$, provided
that
\begin{equation}
  M \geq C \, \mu(\mathbf{\Phi}, \mathbf{W}) \,  K \log N,  
\end{equation}
where $C$ is a constant on the order of unity. This result emphasizes
both the role of the coherence and the potential gain for large images
due to the logarithmic dependence in the pixel size. For incoherent
pairs, we only need on the order of $K \log N$ random samples. 

We have discussed sparse signals above for ease of
exposition. However, the theory extends to approximately sparse
signals and to noisy data. For instance, if the signal is well
approximated by a $K$-sparse signal (some would say that it is
compressible), then the reconstruction error is shown to be
small. Further, the recovery is not sensitive to noise in the sense
that the error degrades gracefully as the signal-to-noise ratio
decreases. We refer to \cite{RIPLESS} and references therein for
quantitative statements.

%%%%%%%%%%%%%%%%%%%%%%%%%%%%%%%%%%%%%%%%%%%%%%%
\section{Compressive Fluorescence Microscopy: Implementation}
%%%%%%%%%%%%%%%%%%%%%%%%%%%%%%%%%%%%%%%%%%%%%%%

\begin{figure}[!h]
\centering
\includegraphics[scale=0.32]{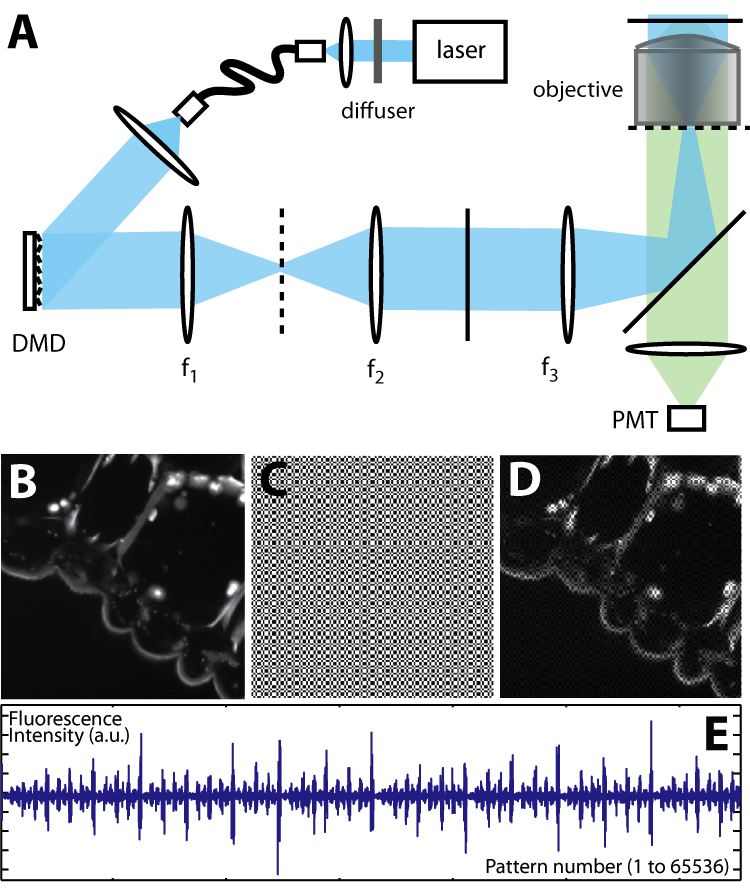}
\caption{{\bf (A) Experimental set-up. The dotted and plain segments
    correspond to planes respectively conjugated to the pupil and
    sample planes. (B) Slice of lily anther (endogenous fluorescence
    with epifluorescence microscopy image recorded on a CCD
    camera). (C) Projection of a Hadamard pattern on a uniform
    fluorescent sample. (D) Projection of the same Hadamard pattern on
    the biological sample. (E) Fluorescence intensity during an
    acquisition sequence.}}
\label{set-up}
\end{figure}

\paragraph{Experimental setup.}
Our setup is based on a standard epifluorescence inverted microscope
(Nikon Ti-E) as shown in Figure \ref{set-up}A. To generate spatially
modulated excitation patterns, we incorporated a Digital Micromirror
Device (DMD) in a conjugate image plane of the excitation path. The
DMD is a 1024-by-768 array of micromirrors (Texas-Instrument Discovery
4100, Vialux, Germany) of size 13.68x13.68 $\mu$m each, and which can
be shifted between two positions oriented at +12$^\mathrm{o}$ or
-12$^\mathrm{o}$ with respect to the DMD surface. They are all
independently configurable at frequency up to 20 kHz. The DMD is
carefully positioned so that the optical axis (defined by the
microscope lens and the dichroic mirror DM (Figure \ref{set-up}A)) is
orthogonal to the plane of the DMD.

As light source, we used a laser bench (Roper Scientific, France)
equipped with two superimposed continuous-wave laser ($\lambda$=488
nm, Coherent, USA and $\lambda$=561 nm, Cobolt, Sweden). The laser
beam first passed through a rotating diffuser or a phase scrambler
(Dyoptika, Ireland) in order to reduce the spatial coherence and was
then coupled to a 200$\mu\text m$ multimode fiber. At the fiber
output, the laser beam was expanded into a 2 cm diameter collimated
beam. This beam was oriented towards the DMD at an angle of incidence
corresponding to twice the tilting angle of the DMD mirrors
(approx.~24$^\mathrm{o}$); a micromirror oriented at +12$^\mathrm{o}$
would reflect the light into the microscope and appear as a bright
pixel in the sample plane and, inversely, micromirrors oriented at
-12$^\mathrm{o}$ appear as dark pixels. Depending on the samples, we
used an air (Nikon, 20X, Plan Apo VC NA 0.75) or an oil-immersion
objective (Nikon, 60X Plan Apo Tirf, NA 1.45). When the 20x air lens
was used, the imaging lenses (lenses $f_1$, $f_2$, $f_3$ in Figure
\ref{set-up}A) were chosen to introduce a 1.5X reduction. The overall
magnification of the image of the DMD on the sample was 1/30 and the
size of a single micromirror equal to 456 nm. When the 60x lens was
used, a different set of imaging lenses was chosen which only served
as a 1x relay; here, the image size of a single micromirror in the
sample plane was equal to 228 nm.  Upon illumination with an intensity
pattern (excitation intensity $\sim$ 20 to 60 W/cm$^2$), the sample
fluorescence was detected on a photomultiplier tube PMT (Hamamatsu)
and sampled at 96kHz using an analog-digital converter board
(PCI-4462, National Instruments, USA) (Figure \ref{set-up}B-D). In CS
measurements, the information on the sample is thus contained in the
variations of the intensity signal as a function of the illumination
pattern (Figure \ref{set-up}E).  The WF image of the sample could also
be directly formed on a camera (ImagEM, Hamamatsu, Japan) placed on
the output port of the microscope. For hyperspectral imaging, the PMT
was replaced by a fast and sensitive spectral detector described later
in the paper. Note that the role of the DMD in our set-up differs from
that in the ``single-pixel camera'' \cite{Baraniuk} or in some other
microscopy setups \cite{Wu2011}.  In the latter, the modulator is
placed between the sample and the detector, meaning that it is used to
select some of the light within the total signal, rather than to
control the excitation pattern. Our choice is motivated by the low
level of fluorescence encountered in biological samples such as living
cells labelled with fluorescent proteins. Indeed, the overall
efficiency of a DMD is 68\% and falls down to 34\% when only one half
of the mirrors are tilted. In our case, the photon collection
efficiency is only limited by the numerical aperture of the microscope
lens and the quantum yield of the detector as in conventional
epifluorescence microscopy.
 
\paragraph{Choice of the illumination patterns.}

\newcommand{\thk}{\tilde{\mathbf{h}}_{\mathbf{k}}} For the practical
implementation of a CS-based image acquisition system, it is essential
to determine which incoherent basis should be used when no prior
information on the signal is available. There are measurement
ensembles, such as the partial Fourier or Hadamard systems, known to
be highly incoherent with the bases in which most natural images are
sparse.  When excitation patterns are generated by micromirrors,
$\phi_k[i]$ is a binary waveform taking on the two values $0$ or
$1$. An appealing choice for $\mathbf{\Phi}$ is then the Hadamard
system known to be incoherent with the Dirac basis and fine scale
wavelets. Since each entry of a Hadamard pattern $\mathbf{h_k}$ is
either $-1$ or $+1$, one defines $\mathbf{\phi_k}$ as a shifted and
rescaled version of $\mathbf{h_k}$ via $\mathbf{\phi_k} =
(\mathbf{h_k}+1)/2$, which can be simply encoded on the DMD.  We used
patterns of size $256\times256$ and $128\times128$ obtained by binning
$2 \times 2$ and $4 \times 4$ groups of micromirrors.  The actual
pattern $\thk$ formed in the sample plane is in fact the convolution
of the ideal pattern $\mathbf{h_k}$ with the point spread function of
the microscope $P_\mathrm{exc}$ in the excitation path.  Figures
\ref{set-up}C and D represent WF images of a Hadamard pattern
projected on a uniform and on a biological sample.  A specificity of
optical imaging is that the sensing elements $\mathbf{\phi_k}$
represent light intensities and are thus nonnegative which, as
discussed later, has important practical implications.

Hadamard waveforms have a sort of spatial frequency (like sinusoids)
which grossly depends on the typical block size of the patterns.  As
the power spectrum of most biological images is generally concentrated
at low frequencies, the flexibility in frequency selection is
important. We introduce two distinct pattern selection strategies
based on the expected spatial content of the sample:
\begin{itemize}
\item When the sample we wish to acquire is sparse in the pixel domain
  as in the case of single molecule or bead imaging, no typical
  frequency range needs to be favored and Hadamard patterns are
  selected uniformly at random.
\item More complex samples have a power spectrum typically decaying
  like a power law. This \textit{a priori} information suggests that
  we should balance low- and high-frequency measurements in order to
  accurately acquire the low-frequency part of the image, which
  accounts for a significant part of the total variance. The {\em
    half-half} strategy then projects the $m/2$ patterns with the
  lowest spatial frequencies to acquire a low-resolution image of the
  sample; the high-resolution content of the image is randomly sampled
  by choosing $m/2$ measurements among the $N - m/2$ remaining
  high-frequency Hadamard patterns. Such an adaptive strategy
  guarantees an accurate determination of the low-frequency content
  while allowing for the estimation of details at a finer scale.
  \end{itemize}

\paragraph{Computational reconstruction.}

In CS, it is essential to enforce the sparsity of the reconstructed
signal in some representation $\mathbf{W}$ that is chosen \textit{a
priori}. The choice of $\mathbf{W}$ highly depends on the spatial
structures of the signal to be reconstructed. One would typically use
a Fourier representation for oscillatory features, wavelets for
point-wise singularities, curvelets for contour-like or filamentary
structures \cite{candes-2004-nfc} and so on. One could also use a
concatenation of all these representations. (If one intends on using
the Fourier basis as a sparsity basis, one would need to scramble the
columns of the Hadamard basis since it would otherwise be coherent
with sinusoids.)

After recording the fluorescence intensity during a sequence of up to
65536 consecutive patterns (Figure \ref{set-up}C), one can
imagine recovering the signal $\mathbf{x}$ from these data by solving the
optimization problem \eqref{program}. Because our measurements are
noisy, it is actually better to relax the constraints into 
% While this is known to be efficient when $x$ is exactly sparse in the
% basis $W$ (\textit{i.e.} $x$ can be exactly represented by $K \ll N$
% non-zero coefficients in $W$), it is generally not optimal when $x$ is
% only approximately sparse. In this more realistic case, all the
% entries of the signal $x$ are non-zero but only a few have a
% significant amplitude. In this case, better reconstruction results can
% be obtained by solving ~(see REF):
\begin{equation}
\label{eq:lasso1}
\min_{\mathbf{x} \in \mathbb{R}^N} \quad \|\mathbf{W}^T \mathbf{x}\|_{\ell_1} \quad \text{subject to} \quad 
\left \| \mathbf{y} - \mathbf{\Phi} \mathbf{x} \right \|_{\ell_2} \leq \epsilon;
\end{equation}
we ask that the fit holds up to the noise level. In the following,
$\mathbf{W}$ will be either an orthonormal basis (\textit{e.g.} Dirac basis) or an overcomplete signal representation (\textit{e.g.} undecimated wavelet frame or curvelet frame). This will be
clearly specified for each individual reconstruction result. For computational reasons, we find
it convenient to solve a relaxed version of this problem, namely,
\begin{equation}
\label{eq:recoptim}
\min_{\mathbf{x} \in \mathbb{R}^N} \quad \|\mathbf{W}^T \mathbf{x}\|_{\ell_1} + \frac{\alpha}{2} \left \| \mathbf{y} - \mathbf{\Phi} \mathbf{x} \right \|_{\ell_2}^2. 
\end{equation}
As is well known, there is a value $\alpha(\epsilon)$ such that the
two programs coincide. For our experiments, we used the NESTA solver
\cite{NESTA} and the regularization parameter $\alpha$ is chosen
empirically depending on the noise level. When the signal is nearly
sparse and the noise level low, it is known that this program finds a
reconstruction with a low mean squared error (MSE).

%%%%%%%%%%%%%%%%%%%%%%%%%%%%%%%%%%%%%%%%%%%%%%%
\section{Sparse Fluorescence Images: Beads, Cells and Tissues}
\label{Exp_results}
%%%%%%%%%%%%%%%%%%%%%%%%%%%%%%%%%%%%%%%%%%%%%%%

\paragraph{Fluorescent beads.}

\begin{figure*}[!t]
\includegraphics[scale=0.150]{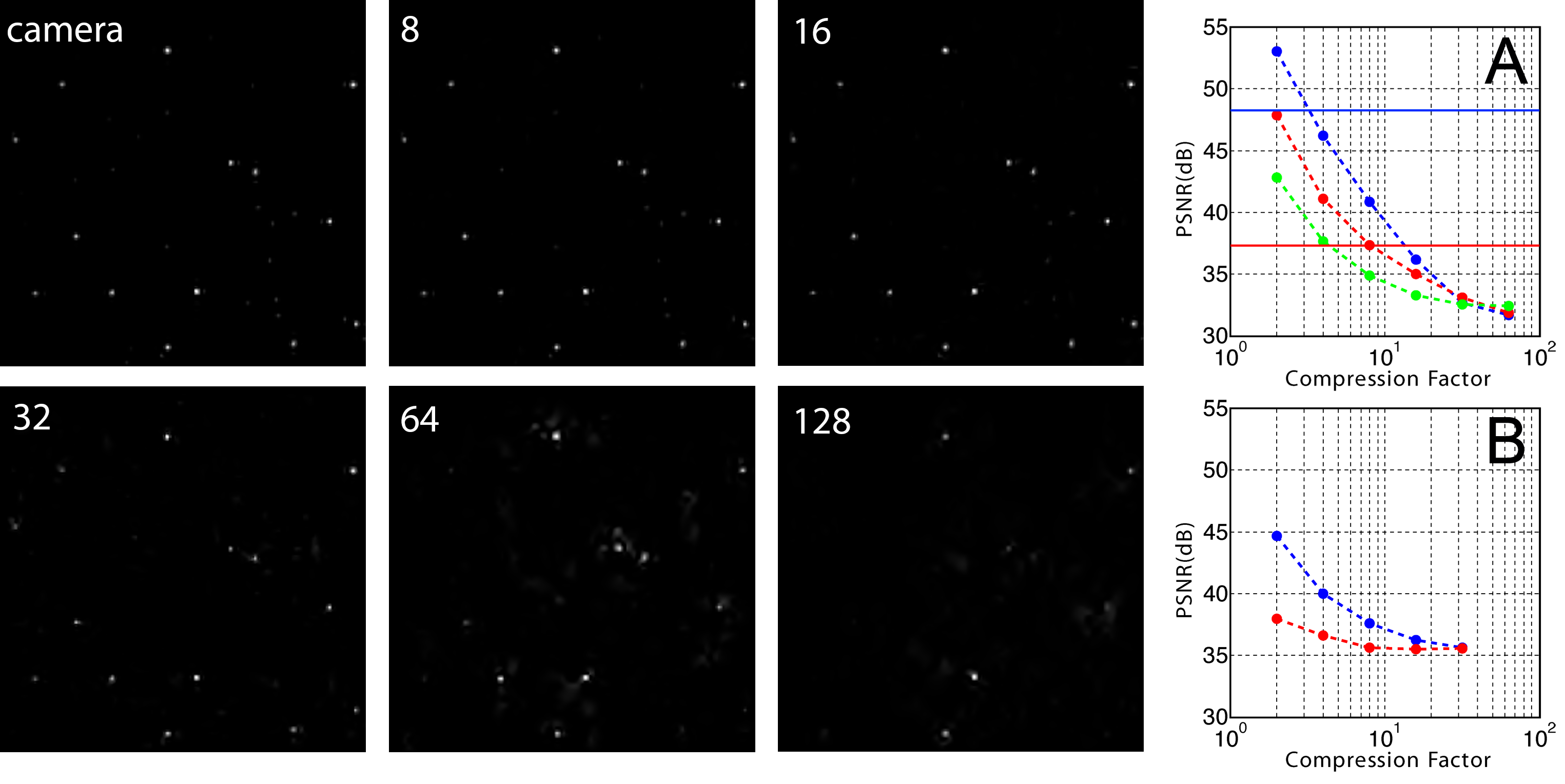}
\centering
\caption{{\bf Top left to bottom right: camera snapshot and
    reconstructed 256-by-256 bead images for values of the
    undersampling ratio equal to 8, 16, 32, 64 and 128. (A) Plot of
    the PSNR (see text) for a nominal illumination level (blue curve)
    and for the same level reduced by a factor 10 (red curve) and a
    factor of 100 (green curve) (simulated data).  The solid lines
    correspond to the PSNR in raster scan for the same surfacic
    illumination. (B) Same as (A) for the experimental data. }}
\label{beads}
\end{figure*}

We first tested our CS microscope (with the 20x objective) on a sample
of fluorescent beads (diameter 2 $\mu$m, peak emission at 520 nm,
Fluorospheres Invitrogen) deposited on a glass coverslip. At a low
density of beads, the WF image is the superposition of a few
fluorescence spots on a dark background, a signal similar to that of
single molecule imaging data in biology \cite{Lord2010}.  As for the
sparsity basis $W$, we obtained nearly equivalent results using the
Dirac basis or a wavelet transform. Here we show images reconstructed
with the wavelet transform and using a number of random $256\times256$
Hadamard patterns decreasing from 16384 down to 512. (To be complete,
we used a weighted $\ell_1$ norm in eq.\,\eqref{eq:recoptim} where the
weight of each coefficient is inversely proportional to scale.)  In
the following, the undersampling ratio is the ratio between the number
$N$ of pixels and the number $M$ of measurements.  As shown in
Figure~\ref{beads}, most of the bead positions are recovered with
undersampling ratios up to 64, corresponding to $M \sim$ 1.5\% of
$N$. At higher undersampling ratios, beads with low intensities are
lost.

To quantify the distorsion of the reconstructed image as a function of
the undersampling ratio, we calculated the Peak Signal-to-Noise Ratio,
$\text{PNSR} = 10\, \log(d^2/\text{MSE})$ where $\text{MSE} = N^{-1}
\|\hat{\mathbf{x}} - \mathbf{x}_{\text{ref}}\|_2^2$, the squared
distance between the reconstructed image from all the $256\times256$
possible measurements and that which only uses a fraction. Above, $d$
is the dynamical range of the reconstruction obtained from a full
sample. As shown in Figure~\ref{beads}A, the PSNR decreases with the
undersampling ratio (blue curve) and seems to reach a plateau at
ratios above 64 where most of the beads are lost. Since beads with low
intensities are lost before brighter beads, we made a second set of
measurements with an excitation light intensity divided by 100 to
assess the effect of illumination on compression efficiency. The red
curve in Figure~\ref{beads}A represents the PNSR of the reconstructed
images as a function of the undersampling ratio. As expected, the PNSR
is lower than that for the nominal illumination and reaches a plateau
at an undersampling ratio of about 10, where almost all the beads are
lost. This clearly shows that the distortion of the reconstructed
image is strongly affected by the amount of detected fluorescent
photons. Indeed at such low intensities, photon noise (also termed,
shot noise) may be significant. 

To further explore the impact of photon noise on the compression
efficiency, we performed numerical simulations on an artificial image
of fluorescent beads made of $50$ Gaussian spots (FWHM 3 pixels)
randomly positioned in the field of view of size $256 \times 256$
pixels. The simulated nominal illumination intensity $I_0$ was set so
that the resulting flux (\textit{i.e.}~the sum of the signals over all
the pixels) was equal to $f_0 = 6.4e3$. Each measurement $y_k$ was
then computed as one realization of a Poisson process with mean
$\langle \mathbf{ \phi_k} , \mathbf{x} \rangle$.

Reconstructions are processed with intensities $I_0$, $I_0/10$ and
$I_0/100$ for a range of undersampling ratios between 2 and 64. As
shown by the PSNR curves (Figure~\ref{beads}B), these simulations
qualitatively reproduce the loss of compression efficiency for
low-light levels but fail to quantitatively estimate the PSNR of the
reconstructed images. This suggests that photon noise is not the only
source of image degradation in our imaging system. A possible additional cause is 
the discrepancies between the theoretical patterns and the effective illumination
profiles in the sample plane.

%In particular, the constrast $(I_{\text{max}}-I_{\text{min}})/(I_{\text{max}} +I_{\text{min}})$ of
% the patterns is lower than 1 for two reasons: first, the reflectivity
% of an ``off'' mirror of the DMD is not zero and can be estimated to be
% about 1\% of that of an ``on'' position. Second, since our
% illumination is made partially incoherent (with a diffuser in the
% excitation path), the modulation transfer function of the microscope
% induces a loss of contrast for higher spatial frequencies. \ejc{[EJC:
 % we still say nothing about the lack of instrument calibration?]}

\paragraph{Lily anther slice.}

In order to investigate the potential of CFM for biological samples,
we imaged slices of endogeneously fluorescent lily anther (Carolina
Biological Supply).  A conventional epifluorescence image of a slice
(Excitation 488nm/Emission 520 nm) recorded on a CCD camera can be
seen in Figure 3 (upper left). The resolution of this image has been
sampled down to $128 \times 128$ pixels. We recorded the same image
with the CFM setup by illuminating the sample with $16384 = 128^2$
different Hadamard patterns (complete basis). For this experiment, we
used a 20x lens with a 0.75 NA. Further, we used curvelets as sparsity
basis $\mathbf{W}$ since they are known to sparsely encode
contour-like structures together with the half-half strategy described
earlier to account for both low-spatial frequencies shapes and higher
frequencies details of the sample structure. (We again used a weighted
$\ell_1$ norm with weights inversely proportional to scale.)
Reconstructed images with varying undersampling ratios from 1 to 8 are
displayed in Figure 3 (top). Here, the method reconstructs images
satisfyingly up to an undersampling ratio of about 8.  Compared to
fluorescent beads, this lower figure can be primarily attributed to
the lesser sparsity.

Another important issue is that this sample is not fully two
dimensional. Due to the thickness of the slice (about 50 $\mu$m in
this case, compared to the focal depth $\sim 1\mu \text m$), the
contrast of Hadamard patterns diminishes away from the focal
plane. This yields a non-modulated background signal and, as further
discussed below, the photon noise associated to this signal affects
the image reconstruction by ``hiding'' the useful information
contained in the intensity fluctuations due to the variations in the
illumination patterns (Figure~1E).

\paragraph{Zyxin-mEOS2 COS7 cells.}

In many biological applications, it is essential to use high
magnification and high NA optics and we thus aimed at testing CFM in
these imaging conditions (oil-immersion objective 60x, NA 1,45). To
overcome the limitations due to the short focal depth of a high NA
objective, we used photoactivation techniques. COS7 cells were
transfected with Zyxin-mEOS2 \cite{Zyxin}.  Zyxin is a protein mainly
expressed in the cellular focal adhesions, at the surface on which the
cells are plated. It was fused to mEOS2, a genetically-encoded
photoconvertible fluorescent protein tag \cite{mEOS} widely used in
super-resolution microscopy, that has green fluorescence in its native
state (Excitation 506nm / Emission 519nm) and can be converted to a
red-emitting state (Excitation 573nm / Emission 584nm) upon
illumination with violet light. The COS7 cells were plated at density
of 100000 cell/ml on 18mm coverslips on a 12 well plate. The cells
were transfected with Eos-Zyxin using chemical transfection (Fugene)
4-5 hours after plating and experiments were performed on live cells
18-30 hours after the transfection. By using an evanescent wave
excitation with a laser at 405 nm, we could convert proteins situated
at the vicinity ($\sim 100$ nm) of the glass coverslip. Therefore, in
our sample, the green emitting fluorophores are located within the 3D
cellular volume while red-emitting proteins constitute a
two-dimensional sample. The superimposed epi-fluorescence images in
the green and red channels (256x256 pixels) are shown in Figure 3
(bottom, left). The same 2D ensemble of photoconverted proteins was
subsequently imaged with the CFM setup with 32768 different 256x256
pixels Hadamard patterns (half of the full basis).  For this set of
data, the pixel size in the sample plane of each Hadamard pattern is
430 nm, about twice the diffraction limit. As for the bead images, we
used a wavelet transform as sparsity basis $\mathbf{W}$.

The reconstructed images for undersampling ratios varying between 2
and 15 are displayed in the second row of Figure 3. For direct
comparison with the conventional epi-fluorescence WF image, a dual
color image obtained by superposition of the red converted Zyxin-mEOS2
image and the green native Zyxin-mEOS2 image is shown in Figure 3. It
is noteworthy that even if the fluorescence emission of transfected
COS7 cells is low compared to the lily anther slice (by about a factor
of 10), the quality of the reconstructed CS images is good for
undersampling ratios up to 8 and starts to be degraded at a ratio of
about 15. The very low background of our two-dimensional sample
clearly enables a better reconstruction. Since fluorescent proteins
are very sensitive to photobleaching, one has to use low illumination
to minimize this effect during acquisition. Thus, photon noise
effects, which manifested themselves only at reduced illumination
intensities in the bead images, appear to be a limiting factor for CS
imaging of less fluorescent and/or photo-damageable samples. The
impact of photon noise on CFM is discussed in more details in a later
section.

\begin{figure*}[htb]
\includegraphics[scale=0.35]{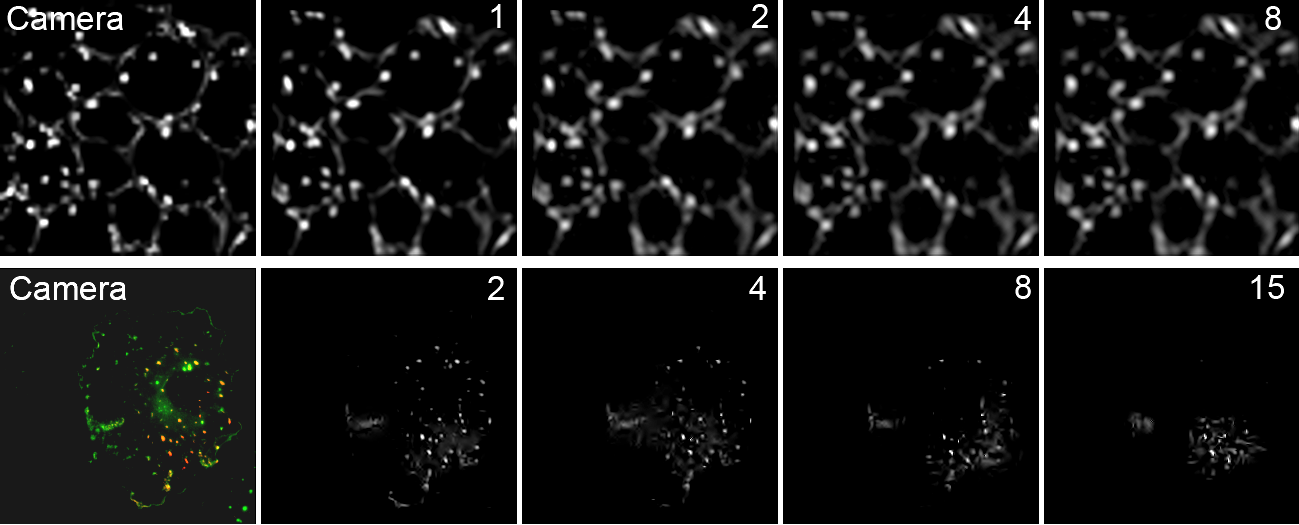}
\centering
\caption{{\bf Upper line: CS imaging of a slice of a lily
    anther. Left: Original image (128x128 pixels) by conventional
    epifluorescence microscopy. Left to right: the same sample imaged
    by CFM with undersampling ratios between 1 and 8. Lower line: CS
    imaging of COS7 cells expressing Zyxin-mEOS2.  Left: superposition
    of the conventional epifluorescence images of the native (green)
    and converted form (red) of the markers. Left to right: CFM images
    of the converted form of the markers at undersampling ratios equal
    to 2, 4, 8 and 15.}}
\label{cells_pollen}
\end{figure*}

\section{Hyperspectral Imaging in CFM}
\label{sec:hyper}

Hyperspectral imaging is defined as the combined acquisition of
spatial and spectral information. In biological imaging, a growing
range of applications such as the study of protein localization and
interactions require quantitative approaches that analyze several
distinct fluorescent molecules at the same time in the same sample
\cite{Hyperspectral Zimmermann}. These applications are in fact
becoming ever more common with the availability of an increasing panel
of fluorescent dyes and proteins with emission ranging from the UV to
the far red \cite{Giepmans}.

Multicolor data are usually acquired by selecting a few distinct
spectral bands. However, in many cases, the incomplete separation of
the different color channels due to the presence of autofluorescence,
along with cross-excitation and emission ``bleed-through'' of one
color channel into the others render the interpretation of multi-band
images difficult and/or ambiguous. To overcome these limitations, it
would often be preferable to record the full spectral information at
each pixel of the image.

\begin{figure}[htb]
\includegraphics[scale=0.155]{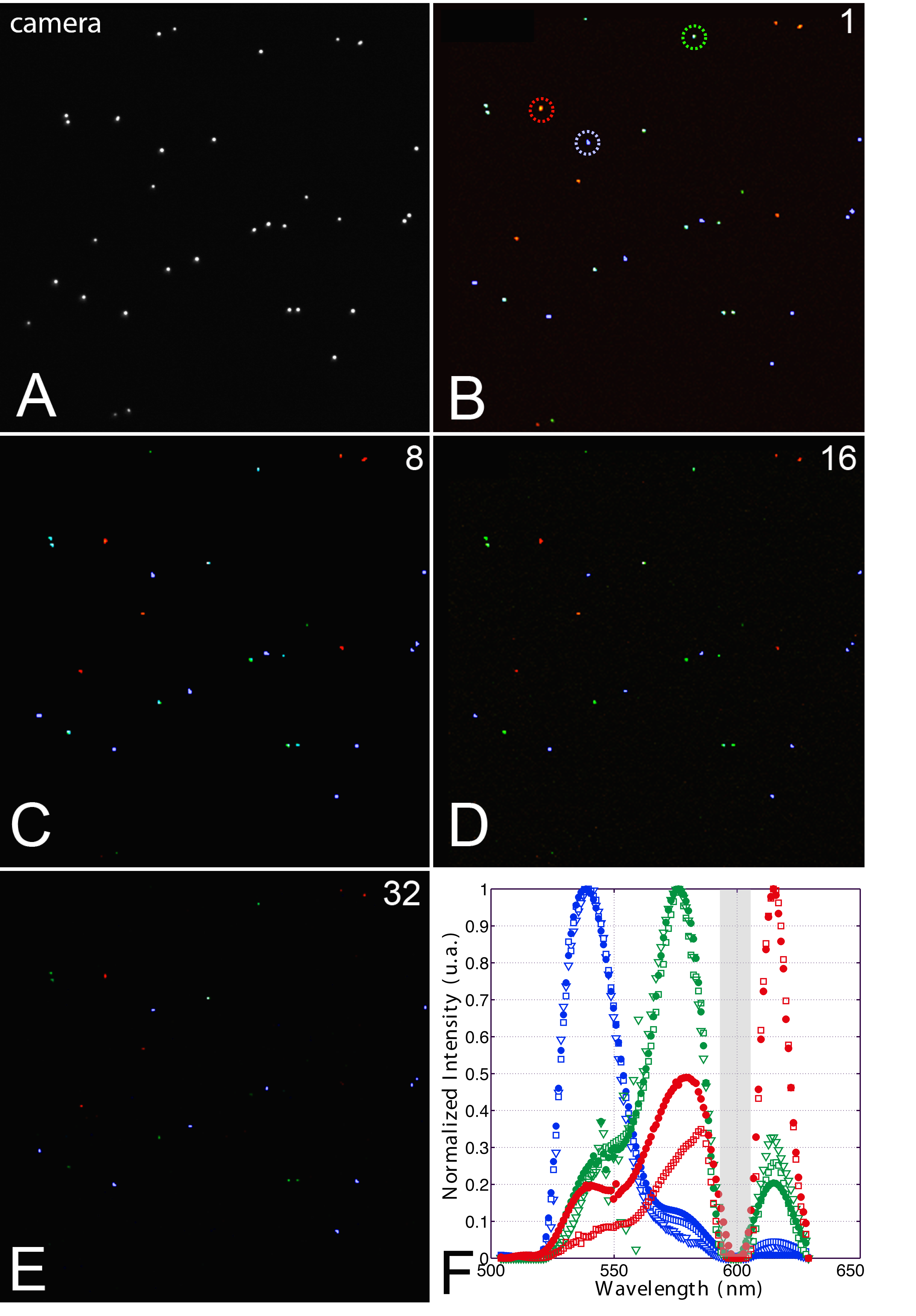}
\caption{{\bf (A)--(E) Camera snapshot and reconstructed 256-by-256
    bead images for undersampling ratios equal to 1, 8, 16 and 32. (F)
    Normalized spectra (128 spectral lines) of 3 individual different
    beads circled in (B) for undersampling ratios equal to 1 (plain
    circles), 32 (squares) and 64 (triangles). The grey area in the
    spectrum represents a rejection band of the dichroic mirror used
    in our setup.}}
\label{multicolor_beads}
\end{figure}

In this context, two elements make the potential benefits of CS particularly appealing for hyperspectral
measurements.  First, a full data acquisition can take up a
very long time since the number of voxels $N$ quickly gets very large. Second, the signal becomes comparably 
sparser as the dimension increases. To demonstrate the possibility of CFM for hyperspectral fluorescent
imaging, we modified the setup and replaced the point detector by a
spectrometer coupled to an EMCCD camera (Evolve 512, Photometrics
USA). The entire spectrum between 520nm and 640nm is recorded on 128x1
pixels. We spin-coated on a glass coverslip a mixture of 3 types of
fluorescent beads (TransFluo Beads, Invitrogen, USA) with different
emission spectra in our detection band (see Figure
\ref{multicolor_beads}A for a gray WF image of the sample). A complete
set of $256 \times 256$ Hadamard patterns was subsequently projected
on the sample and, for each projected pattern, we recorded the
fluorescence spectrum. 

The computational reconstruction of hyperspectral data can be
performed in two different manners. The simplest one is a direct
extension of the monochromatic case and consists in reconstructing
each spectral band independently from the others. This approach,
however, does not fully account for the particular structure of the
hyperspectral data. Rather, it is worthwhile to exploit sparsity in
both the spatial and spectral domains. Hence, we propose a
computational reconstruction by solving the same problem as in
\eqref{eq:recoptim} with the following modification: $\mathbf{x}$ is
now the full 2D--$\lambda$ data cube and $\mathbf{\Phi}$ and
$\mathbf{W}$ are waveforms ${\phi_k}[i,\lambda]$ of both space and
wavelength. In these experiments, $\mathbf{W}$ was obtained by
tensorizing the Dirac basis in space---well adapted to point-wise
structures like beads---with a wavelet basis along the spectral
dimension which is well suited for smooth variations and occasional
transients. A slice of the 2D--$\lambda$ sensing matrix at a fixed
$\lambda$ is the same 2D partial Hadamard transform. 

We obtained full color images by pooling the data cube---see Figure
\ref{multicolor_beads} ---into 3 spectral bands: blue (500-530 nm),
green (530-560 nm) and red (560-630 nm). Such multicolor images are
shown for varying values of the undersampling ratio in Figure
\ref{multicolor_beads}. We observe that almost no degradation is seen
for undersampling factors up to 16.  Furthermore, hyperspectral
reconstructions provide the spectrum of each individual bead from the
reconstructed 2D--$\lambda$ cube (the normalized spectra of 3
different beads are shown in \ref{multicolor_beads}F). The spectra are
correctly reconstructed for undersampling factors up to
64. Interestingly, when reducing the number of measurements, the
distortion primarily affects the low intensity parts of the spectra,
similar to the effect of increasing undersampling on the dimmer beads
in monochromatic images.

 \section{Discussion and Perspectives}
\label{sec:discussion}

We have developed an imaging approach based on the concepts of CS
theory which, on samples relevant for biological imaging, allows the
reconstruction of fluorescence images with undersampling ratios up to
64. While this constitutes a significant gain over undersampling
ratios achieved in prior CS-based imaging approaches, several factors
(computational, instrumental or noise-related) still contribute to
limit the current performances of CFM. Below, we discuss them as well
as the prospects for future developments and applications.

\paragraph{Point spread function and its modeling.}

In the current implementation of CFM, we neglected the role of the
point spread function (PSF) of the excitation pathway (corresponding
to the lenses $f_1, f_2$ and $f_3$ and the objective) and used the
idealized matrix with zeros and ones (depending whether a pixel is
illuminated or not) for the inversion. In reality, the PSF acts as a
low pass filter, meaning that those patterns illuminating the sample
are spatially smoother than their theoretical counterparts. In the
framework of CS, it is well known that recovering a compressed signal
from a sensing matrix that departs from the real one may, in general,
dramatically degrade the reconstruction quality. In our case, however,
this approximation only has minor consequences. This holds for
essentially two reasons. In our experiments, the individual pixel size
of the Hadamard patterns was at least twice as large as the
diffraction-limit of the microscope. In other words, the PSF only has
a minor filtering impact on the patterns.  Second, neglecting the
effect of the PSF in the reconstruction does not dramatically reduce
the quality of reconstruction. To see this, let $\mathbf{f}$ be the
PSF of the lens so that the collected fluorescence is of the form $y_k
= \<\mathbf{\phi_k} * \mathbf{f}, \mathbf{x}\>$ (we measure the dot
product between the object of interest and the theoretical patterns
convolved with the PSF). If the PSF is symmetric around the origin, we
have
$$
\mathbf{y} = \mathbf{\Phi} \mathbf{F} \mathbf{x}
$$
in which $\mathbf{F}$ is the linear convolution with the PSF.
Hence, neglecting the PSF recovers a signal $\tilde{\bx}$ obeying $\by
= \bPhi \tilde{\bx}$; in other words, one gets $\tilde{\bx} =
\mathbf{F} \bx$ which is $\bx$ at the resolution of the
microscope. Indeed, assuming that the lenses the excitation
and emission PSFs are identical, the CSM has the same resolution as its wide-field
equivalent. Now if one wishes to account for the PSF, one would need
to solve a joint decompression--deconvolution problem. If the
deconvolution should provide a higher resolution image, it is
well-known that it is usually at the cost of noise amplification. Here
again, the sparsity prior used for decompression should of course help
regularizing this deconvolution step.

\paragraph{Noise and MSE.} It is crucial to examine noise figures in
our setup. There are two important facts affecting the quality of
reconstruction: the noise distribution associated with CS-type data
and the undersampling ratio.

We assume below that photonic noise is the limiting source of noise,
an assumption which sets CS-based optical systems apart from the
common theoretical CS framework \cite{Willet11b,Baraniuk}. To
understand the important trade-offs, we compare the respective
situations of RS and CS, two point detection imaging
techniques. Below, we put $\IRS$ and $\ICS$ for the excitation
intensity per unit area and per unit time during RS and CS
acquisition. Likewise, $\TRS$ and $\TCS$ are durations of excitation
(for a single measurement). Finally, we set $\lRS = \IRS \times \TRS$
and similarly for $\lCS$. In practice, all these parameter values are
adjusted according to factors that one wants to optimize (acquisition
speed, sensitivity, photobleaching and so on) and they must be
evaluated on a case by case basis. Hence, rather than an exhaustive
comparison of the relative merits of RS and CS acquisition, the
discussion below aims at providing a general framework to understand
the nature of the noise in CS measurements.

In the case of RS, the $i$th pixel measurement is distributed as a
Poisson random variable with mean and variance $\lRS \times x[i]$.
Using the scaled observed data yields a per-pixel MSE equal to

%\begin{eqnarray}
\[
\text{MSE(RS)} = N^{-1} \sum_i \mathbb{E} (\hat{x}[i] - x[i])^2 =
\lRS^{-1} \times \bar x, 
% \mathrm{with} \qquad \bar x &=& N^{-1} \sum_i x[i].
%\end{eqnarray}
\]
where $\bar x = N^{-1} \sum_i x[i]$. For CS, suppose first that we
collect all Hadamard measurements (no undersampling).  Each
measurement is an independent Poisson variable with mean $\lCS \times
\< \mathbf{\phi_k}, \mathbf{x}\>$. With patterns of the form $\frac12
(1+\mathbf{h_k})$ where $\mathbf{h_k}$ is a Hadamard sequence, one can
decompose the mean value of $y_k$, as
\[
\lCS \times\< \mathbf{\phi_k}, \mathbf{x}\> = \frac{\lCS}{2} [N \bar x
+ \<\mathbf{h_k}, \mathbf{x}\>].
\]
Hence, this is the sum of a DC offset and a Fourier-like
component. The presence of the DC offset (which prevents the optical
patterns from being negative) impacts the data SNR. Indeed, a possible
source of concern is that for many high-frequency components,
$N\bar{x}$ may be much greater than the magnitude of $ \<
\mathbf{h_k}, \mathbf{x}\>$---the DC component dominates the
high-frequency coefficients. Therefore, when measuring a
high-frequency component, we need to deal with a large amount of noise
coming from the average fluorescence of the sample under study.  This
situation is arguably very different than in other applications---for
instance, the acquisition of radio-frequency signals---where one can
use sensing waveforms that take on negative values by switching the
phase of the object we wish to acquire \cite{RMPI}. Inverting the
Hadamard matrix gives a noisy image $\hat x[i]$ obeying
\[
\mathbb{E} \, \hat x[i] = x[i] \quad \text{ and } \quad
\text{Var}(\hat x[i]) = \frac{2}{\lCS} \bar x
\]
(see the Appendix for details). 
In contrast to RS microscopy, we see that CFM yields a spatially
invariant noise level in the pixel domain. In other words, by
measuring the image projection on an incoherent basis, the noise gets
spread equally over all the pixels. Hence, before applying any
processing, the CS situation is more favorable for recovering brighter
areas but less so for dimmer regions. Summing up gives 
\[
\text{MSE(CS)} =  2{\lCS}^{-1} \times \bar x. 
\]
This analysis also shows that it is essential to minimize the sources
of signal which could contribute to a constant background and increase
$\bar x$. This offset could be due to the nature of the sample itself
but can also originate from stray light or out-of-focus
fluorescence. These considerations explain, at least qualitatively,
why the {\em beads} sample, where the background $\bar x$ is low
compared to the bright spots and is purely bi-dimensional, is
favorable for CFM.

To consider the effect of statistical estimation procedures or data
processing, consider the \textit{beads} sample again in which sparsely
distributed beads are located in the field of view. Each bead is an
isolated bright spot surrounded by wide non-fluorescent areas. Suppose
then that we were to apply a thresholding estimator, setting to zero
all intensities below a certain level, and keeping those above
threshold. Then one would obtain a very low MSE in the CS setting
since dark pixels would be correctly set to zero while bright pixels
would have a variance that is orders of magnitude lower than that
achievable in the RS case. In short, thresholding would effectively
filter out the off-support noise and the situation would be extremely
favorable to the CS approach. Quantitatively, if there are $K$ bright
pixels, the error after estimation for RS and CS would behave like
\begin{equation}
\label{eq:CS1}
\text{MSE(RS)} = {\lRS}^{-1} \times \bar x, \qquad \text{MSE(CS)} = 2
{\lCS}^{-1} \times \frac{K}{N} \times \bar x.
\end{equation}
Note the potentially enormous reduction in MSE by the factor $K/N$.
Conversely, RS would be more effective for smooth and bright images
(sparse images in the frequency domain).

The comparisons between RS and CS microscopy above are valid as long
as the same illumination intensity per pixel is assumed and the number
of measurements $M$ is equal to the number of pixels. We now discuss
the effect of undersampling. Now the CS recovery is both an inversion
and a denoising algorithm and the recovery error depends on the
compressibility of the signal (on how sparse it is).  For instance, in
the fully sampled case where sparsity is assumed in the spatial
domain, the CS recovery would essentially invert the Hadamard matrix
and then apply soft-thresholding to the output. Now suppose for
simplicity that the signal is $K$-sparse and that the number $M$ of
measurements is sufficient for perfect recovery from noiseless
data. Then the squared recovery error from $M$ noisy measurements as
above would roughly scale like
\begin{equation}
  \label{eq:CS2}
  \text{MSE(CS)} = C_0 \times {\lCS}^{-1} \times \frac{K}{M} \times \bar x,
\end{equation}
where $C_0$ is a small numerical constant (typically on the order of unity).
Hence, the main difference with the MSE available from a full
sample is a loss of a factor $N/M$---the undersampling ratio---in the
MSE, compare \eqref{eq:CS1}. In other words, halving the number of
samples---everything else, namely, intensity and duration of
excitation remaining the same---doubles the MSE. The same conclusion
applies for approximately sparse signals for which the variance
component of the MSE dominates the squared bias.  Here again, the more
compressible the signal (the smaller $K$), the better the performance
(e.~g.~the {\it beads} sample is favorable to CFM). We would like to
also note that $K/M$ is still much smaller than one so that even
though we are sampling less, we may still end up with a much better
MSE than in RS.

To summarize, CFM is effective when:  1) the most informative parts of
the sample are brighter than its mean value and 2) the sample is
highly compressible. Notice that while these two requirements are
sample dependent, the second gives some flexibility since one can
select a representation in which a class of signals has an optimally
sparse representation.

% In fact, assuming that the limiting noise is photonic noise, each CS measurement is an independent Poisson variable with mean $\ICS \times \< \mathbf{\phi_k}$ where $\ICS$ is the pixelwise intensity of the illumination. Then, in can be shown - see in the Appendix~\ref{sec:SNR} - that the SNR of the CS measurements is equal $\rho_{\text{CS}} = \sqrt{\ICS \, [N \bar x + \<h_k, x\>]/2}$ where $\bar x$ stands for the mean value of the sample. In this expression, the term $N \bar x$ is directly related to the non-negativity of the sensing matrix. Interestingly, the SNR is generally dominated by offset-related term as  $N \bar x \gg  |\<h_k, x\>|$ in general.\\
% As a comparison, the SNR of the each raster scan measurements is
% equal to $\sqrt{\ICS x[i]}$ at pixel $i$. Assuming that the number
% of measurements is equal to $N$ ({\it i.e.} there is no
% compression), we show in the Appendix that the SNR of the signal
% acquired by CS at pixel $i$ is approximately equal to $\sqrt{\ICS
%   \bar x}$.

\paragraph{Impact of the sample thickness. }

One important issue in CFM has to do with the patterned illumination
of thick fluorescent samples. Indeed, with a wide-field linear
excitation, the illumination propagates throughout the sample and
causes the entire volume to fluoresce. Since the optical transfer
function of a circular aperture (such as a microscope lens) has a
bandwidth that decreases with defocusing \cite{Mertz}, the contrast of
the pattern diminishes away from the focal plane. As a result, the
fluorescence coming from out-of-focus planes is not modulated as a
function of the patterns. In fact, this property serves as basis for
optical sectioning in structured illumination microscopy (SIM)
\cite{Review SIM Mertz}. In the case of CFM, the off-focus signal
contributes to an offset signal on the detector, which, as explained
above, tends to significantly degrade the quality of the
reconstruction. A few strategies can be considered to add sectioning
capabilities to CS based imaging systems \cite{Prather09_3},
\cite{Prather09_2} \cite{Wu2011}. One recently demonstrated approach
is based on the rejection of the off-focus signal, in a way similar to
that of a programmable array microscope \cite{Wu2011}. Another method
is to avoid generating any off-focus signal at all. We demonstrated it
for photoactivation using an evanescent-wave (Figure
\ref{cells_pollen}) and this can be extended to 2D activation within
  the sample volume with two-photon temporal focusing activation
  \cite{Temporal Focusing} or light sheet illumination  \cite{Huiskens2004}. In the
  long-term, an even better strategy is to illuminate with an
  incoherent basis of 3D patterns and, subsequently, to directly
  reconstruct the sample in 3D.

\section{Conclusions and Prospect}

This paper presented the principles and implementation of compressive
sensing in fluorescence microscopy together with its applications in
biomedical imaging. Our approach, which is based on a patterned excitation of the sample combined with a point-detection
of the emitted fluorescence, readily allows for substantial undersampling
gains when compared to traditional raster-scanning approaches. It could also be useful in situations, such as a
diffusing media, where direct imaging on a multi-pixel detector is not possible.  Furthermore, we have set
forth a distinctive prospect for hyperspectral acquisition, which has great potential for multicolor single molecule imaging. 
More generally, the acquisition of 3D, 4D (three spatial dimensions and one spectral
or temporal dimension) or even higher dimensional signals puts
unrealistic constraints on system resources. It is indeed hard to
imagine that one would want to sample such huge data cubes at rates
anywhere close to the Shannon rate. The key is that multidimensional
signals become increasingly redundant in the sense that their
information content grows at a much lower rate than the number of
voxels. For example, movies are comparably far more compressible than
still pictures. Likewise, hyperspectral movies are far more redundant
than monochromatic movies, and so on.  Expressed differently, the
ratio between the number of degrees of freedom and the number of
voxels decreases very rapidly as the dimension increases.  The extreme
sparsity of higher dimensional signals cannot be ignored and we expect
the advantages of CFM to become paramount in such
applications.

\appendix

\newcommand{\bH}{\mathbf{H}} \newcommand{\bS}{\mathbf{S}}
\newcommand{\bone}{\mathbf{1}} \newcommand{\bd}{\mathbf{d}}
\newcommand{\beone}{\mathbf{e}_1} \newcommand{\E}{\mathbb{E}}

\noindent This short appendix justifies our SNR calculations, and we begin by
introducing some notation. We denote by $\bH$ the Hadamard matrix and
by $\bone$ the vector with all entries equal to one.  Hence, acquiring
all Hadamard patterns gives us independent Poisson variables with
means
\[
\bd = \bS \bx := \frac{1}{2} (\bone \bone^T + \bH) \bx. 
\]
Here, we set $\lCS = 1$ as the general case can be obtained via a
simple rescaling.  Hence, our estimate is of the form
\[
\hat{\bx} = \bS^{-1} \by,   
\]
and it is easy to verify that 
\[
\bS^{-1} = \frac{2}{N}\Bigl(\bH^T  - \frac{N}{2} \beone \beone^T\Bigr) 
\]
in which $\beone = (1, 0, \ldots, 0)$. (Observe that the first entry
is a bit special here---a pixel at this location is always
illuminated---and that we could always shift the Hadamard matrix as to
select any {\em special} pixel.) Since $\E \by = \bd = \bS \bx$, we
have $\E \hat{\bx} = \bx$. Further,
\[
\text{Cov}(\hat{\bx}) = \bS^{-1} \text{Cov}(\by) \bS^{-T},
\]
where $\text{Cov}(\hat{\bx})$ is the covariance matrix of the random
vector $\hat{\bx}$. Since $\text{Cov}(\by)$ is the diagonal matrix
with entries $\bd = (d_1, d_2, \ldots, d_N)$, we have
\[
\text{Var}(\hat{x}[i]) = \sum_{j = 1}^N |S^{-1}[i,j]|^2 \, d_j.
\]
To proceed, one verifies that 
\[
\frac{N}{2} \, |S^{-1}[i,j]| = \begin{cases} \frac{N}{2} -1, & (i,j) = (1,1), \\
  1, & (i,j) \neq (1,1),
\end{cases}
\]
and that $\sum_{j=1}^N d_j = \frac12 (N^2 \bar x + N x[1])$ together
with $d_1 = N\bar x$. Plugging in gives
\[
\text{Var}(\hat{x}[i])  = \begin{cases} 
(N - 2) \bar x + \frac{2}{N} x[1],  & i = 1,\\
  2\bar x + \frac{2}{N} x[1], & i \neq 1
\end{cases}
\]
(again this highlights the role played by the special pixel).  Now
suppose that the special pixel is chosen so that there is no probe at
this location (which can always be arranged). Then we would know that
$\bx[1] = 0$ (and would not bother estimating the density at that
location) and for $i \neq 1$, we would have
\[
\text{Var}(\hat{x}[i]) = 2\bar x
\]
as claimed.

\begin{acknowledgments}

  Deepak Nair is gratefully acknowledged for providing the Zyxin-mEOS2
  transfected COS7 cells and for his careful reading of the
  manuscript. This work was supported by a grant from the EADS
  foundation. M.C. is supported by the EADS foundation and the conseil
  r\'egional Aquitaine. E.~C.~is partially supported by NSF via grant
  CCF-0963835 and the 2006 Waterman Award, by AFOSR under grant
  FA9550-09-1-0643 and by ONR under grant N00014-09-1-0258.
\end{acknowledgments}

%% PNAS does not support submission of supporting .tex files such as BibTeX.
%% Instead all references must be included in the article .tex document. 
%% If you currently use BibTeX, your bibliography is formed because the 
%% command \verb+\bibliography{}+ brings the <filename>.bbl file into your
%% .tex document. To conform to PNAS requirements, copy the reference listings
%% from your .bbl file and add them to the article .tex file, using the
%% bibliography environment described above.  

%%  Contact pnas@nas.edu if you need assistance with your
%%  bibliography.

% Sample bibliography item in PNAS format:
%% \bibitem{in-text reference} comma-separated author names up to 5,
%% for more than 5 authors use first author last name et al. (year published)
%% article title  {\it Journal Name} volume #: start page-end page.
%% ie,
% \bibitem{Neuhaus} Neuhaus J-M, Sitcher L, Meins F, Jr, Boller T (1991) 
% A short C-terminal sequence is necessary and sufficient for the
% targeting of chitinases to the plant vacuole. 
% {\it Proc Natl Acad Sci USA} 88:10362-10366.

%% Enter the largest bibliography number in the facing curly brackets
%% following \begin{thebibliography}

\end{article}
%%%%%%%%%%%%%%%%%%%%%%%%%%%%%%%%%%%%%%%%%%%%%%%%%%%%%%%%%%%%%%%%

%% Adding Figure and Table References
%% Be sure to add figures and tables after \end{article}
%% and before \end{document}

%% For figures, put the caption below the illustration.
%%
%% \begin{figure}
%% \caption{Almost Sharp Front}\label{afoto}
%% \end{figure}

%% For Tables, put caption above table
%%
%% Table caption should start with a capital letter, continue with lower case
%% and not have a period at the end
%% Using @{\vrule height ?? depth ?? width0pt} in the tabular preamble will
%% keep that much space between every line in the table.

%% \begin{table}
%% \caption{Repeat length of longer allele by age of onset class}
%% \begin{tabular}{@{\vrule height 10.5pt depth4pt  width0pt}lrcccc}
%% table text
%% \end{tabular}
%% \end{table}

%% For two column figures and tables, use the following:

%% \begin{figure*}
%% \caption{Almost Sharp Front}\label{afoto}
%% \end{figure*}

%% \begin{table*}
%% \caption{Repeat length of longer allele by age of onset class}
%% \begin{tabular}{ccc}
%% table text
%% \end{tabular}
%% \end{table*}

\end{document}